\def\bi{\begin{Keywords}}
\def\ei{\end{Keywords}}
\def\be{\begin{equation}}
\def\ee{\end{equation}}
\def\ba{\begin{array}}
\def\ea{\end{array}}

\documentclass[aps,amsmath,amssymb,amsfonts]{revtex4}
\usepackage{graphicx}
\usepackage{epstopdf}
\def\qed{\leavevmode\unskip\penalty9999 \hbox{}\nobreak\hfill
     \quad\hbox{\leavevmode  \hbox to.77778em{%
               \hfil\vrule   \vbox to.675em%
               {\hrule width.6em\vfil\hrule}\vrule\hfil}}
     \par\vskip3pt}

\newtheorem{theorem}{Theorem}

\begin{document}
\title{ Tighter constraints of multiqubit entanglement in terms of R\'{e}nyi-$\alpha$ entropy}
\author{Meng-Li Guo$^{1}$}
\author{Bo-Li$^{2}$}
\email{libobeijing2008@163.com.}
\author{Zhi-Xi Wang$^3$}
\author{Shao-Ming Fei$^{3,4}$}

\affiliation{$^1$Department of Mathematics, East China University of Technology, Nanchang 330013, China\\
$^2$School of Mathematics and Computer science, Shangrao Normal University, Shangrao 334001, China\\
$^3$School of Mathematical Sciences, Capital Normal University, Beijing 100048, China\\
$^4$Max-Planck-Institute for Mathematics in the Sciences, 04103, Leipzig, Germany}

\begin{abstract}
Quantum entanglement plays essential roles in quantum information processing.
The monogamy and polygamy relations characterize the entanglement distributions
in the multipartite systems. We present a class of monogamy inequalities related to the $\mu$th power of the entanglement measure based on R\'{e}nyi-$\alpha$ entropy, as well as polygamy relations in terms of the $\mu$th powered of R\'{e}nyi-$\alpha$ entanglement of assistance.
These monogamy and polygamy relations are shown to be tighter than the existing ones.
\end{abstract}

\maketitle


\section{Introduction}
Quantum entanglement is one of the most quintessential features of quantum mechanics, which distinguishes the
quantum from the classical world and plays essential roles in quantum information processing \cite{c1,c2,c3,Bo,Ho},
revealing the basic understanding of the nature of quantum correlations.
One distinct property of quantum entanglement is that a quantum system entangled
with another system limits its sharing with other systems,
known as the monogamy of entanglement \cite{c4,c5}.
The monogamy of entanglement can be used as a resource to distribute a secret key
which is secure against unauthorized parties \cite{c6,c7}.
It also plays a significant role in many field of physics
such as foundations of quantum mechanics \cite{c8,c9},
condensed matter physics \cite{c11}, statistical mechanics \cite{c8},
and even black-hole physics \cite{c12,c13}.

The monogamy inequality was first introduced by Coffman-Kundu-Wootters(CKW),
by using tangle as a bipartite entanglement measure in three-qubit systems \cite{c14}, and then generalized to multiqubit systems based on various entanglement measure \cite{c15}.
The assisted entanglement is a dual concept to bipartite entanglement measure,
which shows polygamy relations in multiparty quantum systems.
For a three-qubit state $\rho_{ABC}$, a polygamy inequality was introduced as \cite{c17}
$\tau^a(\rho_{A|BC})\leq\tau^a (\rho_{A|B})+\tau^a(\rho_{A|C})$,
where $\tau^a(\rho_{A|BC}) = \max \sum_{i}p_{i} \tau(|\psi_{i}\rangle_{A|B})$ is the tangle of assistance \cite{c16,c17},
with the maximum taking over all possible pure state decompositions of
${\rho _{AB} }= \sum_{i} {p_i \left| {\psi _i } \right\rangle _{AB} \left\langle {\psi _i } \right|}$.
This tangle-based polygamy inequality was extended to multiqubit systems
and also high-dimensional quantum systems in terms of various entropy entanglement measures \cite{c18,c19}.
General polygamy inequalities of entanglement is also established in arbitrary dimensional multipartite quantum systems \cite{c20,c21,Luo,zhu,jin3,JSK7}.

In this paper, we investigate the
monogamy and polygamy constraints based on the $\mu$th power of entanglement measures in terms of the R\'{e}nyi-$\alpha$ entropy for multiqubit systems.
By using the Hamming weight of binary vectors we present a class of monogamy inequalities for multiqubit entanglement based on the $\mu$th power of R\'{e}nyi-$\alpha$ entanglement (R$\alpha$E)  \cite{c22} for $\mu \geq 1$.
For $0 \leq \mu \leq 1$, we introduce a class of tight polygamy inequalities
based on the $\mu$th power of the R\'{e}nyi-$\alpha$ entanglement of assistance (R$\alpha$EoA).
Then, we show that both the monogamy inequalities with $\mu \geq1$ and the
polygamy inequalities with $0\leq \mu \leq 1$ can be further improved to be tighter under certain conditions. These monogamy and polygamy relations are shown to be tighter than the existing ones.
Moreover, our monogamy inequality is shown to be more effective for the counterexamples
of the CKW monogamy inequality in higher-dimensional systems.

\section{Preliminaries}
We first recall the conceptions of R\'{e}nyi-$\alpha$ entropy, R\'{e}nyi-$\alpha$ entanglement, and multiqubit monogamy and polygamy inequalities.
For any $\alpha >0$, $\alpha \neq 1$, the
R\'{e}nyi-$\alpha$ entropy of a quantum state $\rho$ is defined as \cite{c23}
\begin{equation*}
S_\alpha(\rho) = \frac{1}{1-\alpha}\log (\mbox{tr} \rho^\alpha).
\end{equation*}
$S_{\alpha}(\rho)$ reduces to the von Neumann entropy when $\alpha$ approach to 1.

The R\'{e}nyi-$\alpha$ entanglement (R$\alpha$E) $E_{\alpha}\left(|\psi \rangle_{AB} \right)$ of a bipartite pure state $|\psi \rangle_{AB}$ is defined as
\begin{equation*}
E_{\alpha}\left(|\psi \rangle_{AB} \right)=S_{\alpha}(\rho_A),
\label{pure}
\end{equation*}
where $\rho_A= \mathrm{Tr}_B |\psi \rangle_{AB} \langle \psi|$ is the reduced state of system $A$.
For a mixed state $\rho_{AB}$, the R\'{e}nyi-$\alpha$ entanglement is given by
\begin{equation*}
E_{\alpha}\left(\rho_{A|B} \right)=\min \sum_i p_i E_{\alpha}(|\psi_i \rangle_{A|B}),
\end{equation*}
where the minimum is taken over all possible pure state decompositions of
$\rho_{AB}=\sum_{i}p_i |\psi_i \rangle_{AB} \langle \psi _i|$.

As a dual concept to R$\alpha$E, the
R\'{e}nyi-$\alpha$ entanglement of assistance (R$\alpha$EoA) is introduced as
\begin{equation}\label{EoA}
E^{a}_{\alpha}\left(\rho_{A|B} \right)=\max \sum_i p_i E_{\alpha}(|\psi_i \rangle_{A|B}),
\end{equation}
where the maximum is taken over all possible pure state decompositions of $\rho_{AB}$ \cite{c24}.

For any multiqubit state $\rho_{A B_0 \cdots B_{N-1}}$,
a monogamous inequality has been presented in Ref. \cite{c24} for $\alpha \geq 2$,
\begin{equation}
E_{\alpha}\left( \rho_{A|B_0 \cdots B_ {N-1}}\right)\geq \sum_{i=0}^{N-1}E_{\alpha}\left(\rho_{A|B_i}\right),
\label{n7}
\end{equation}
where $E_{\alpha} ( \rho_{A|B_0 \cdots B_{N-1}})$ is the
R$\alpha$E of $\rho_{A B_0 \cdots B_N-1}$ with respect to the bipartition between $A$ and $B_0 \cdots B_{N-1}$, and $E_{\alpha}\left(\rho_{A| B_i}\right)$ is the R$\alpha$E of the reduced density matrix
$\rho_{A B_i}$, $i=0,\cdots,N-1$.

In addition, a class of polygamy inequalities has been obtained for multiqubit systems,
\begin{equation}
E^{a}_{\alpha}\left( \rho_{A|B_0 \cdots B_ {N-1}}\right)\leq \sum_{i=0}^{N-1}E^{a}_{\alpha}\left(\rho_{A|B_i}\right),
\label{n8}
\end{equation}
for $0 \leq \alpha \leq 2$, $\alpha \neq 1$, where $E_{\alpha} ( \rho_{A B_0 \cdots B_{N-1}})$
is the R$\alpha$EoA of $\rho_{A B_0 \cdots B_N-1}$ with
respect to the bipartition between $A$ and $B_0 \cdots B_{N-1}$, and $E^{a}_{\alpha}\left(\rho_{A| B_i}\right)$ is the R$\alpha$EoA of the reduced density matrix $\rho_{A B_i}$, $i=0,\cdots,N-1$.

In Ref. \cite{c25}, Kim established a class of tight monogamy inequalities of multiqubit entanglement in terms of Hamming weight. For any nonnegative integer $j$ with binary expansion
$j=\sum_{i=0}^{n-1} j_i 2^i$, where $\log_{2}j \leq n$ and $j_i \in \{0, 1\}$ for $i=0, \cdots, n-1$,
one can always define a unique binary vector associated with $j$,
$\overrightarrow{j}=\left(j_0,~j_1,~\cdots ,j_{n-1}\right)$.
The Hamming weight $\omega_{H}\left(\overrightarrow{j}\right)$ of the binary vector $\overrightarrow{j}$
is defined to be the number of $1's$ in its coordinates \cite{c26}.
Moreover, the Hamming weight $\omega_{H}\left(\overrightarrow{j}\right)$ is bounded above by $\log_{2}j$,
\begin{equation}\label{n11}
\omega_{H}\left(\overrightarrow{j}\right)\leq \log_{2}j \leq j.
\end{equation}

Kim proposed the tight constraints of multiqubit entanglement based on Hamming weights \cite{c25},
\begin{equation}\label{n12}
[E_{\alpha}(\rho_{A|B_0B_1\ldots B_{N-1}})]^\mu \geq \sum\limits_{j=0}^{N-1} \mu^{\omega_{H}(\vec{j})}[E_{\alpha}(\rho_{A|B_j})]^\mu
\end{equation}
for $\mu \geq 1$, and
\begin{equation}\label{n13}
[E_{\alpha}^a(\rho_{A|B_0B_1\ldots B_{N-1}})]^\mu \leq \sum\limits_{j=0}^{N-1}\mu^{\omega_{H}(\vec{j})}[E_{\alpha}^a(\rho_{A|B_j})]^\mu
\end{equation}
for $0\leq\mu \leq 1$.
Inequalities \eqref{n12} and \eqref{n13} are then further written as
\begin{equation*}
[E_{\alpha}(\rho_{A|B_0B_1\ldots B_{N-1}})]^\mu \geq \sum\limits_{j=0}^{N-1} \mu ^{j}[E_{\alpha}(\rho_{A|B_j})]^\mu
\end{equation*}
for $\mu \geq 1$, and
\begin{equation*}
[E_{\alpha}^a(\rho_{A|B_0B_1\ldots B_{N-1}})]^\mu \leq \sum\limits_{j=0}^{N-1}\mu ^{j}[E_{\alpha}^a(\rho_{A|B_j})]^\mu
\end{equation*}
for $0 \leq \mu \leq 1$.

In the following we show that these inequalities above can be further improved to be much tighter under certain conditions, which provide tighter constraints on the multiqubit entanglement distribution.

\section{Tighter constraints of multiqubit entanglement in terms of R$\alpha$E}

We first present a class of tighter monogamy and polygamy inequalities of multiqubit entanglement
in terms of the $\mu$th power of R$\alpha$E.
We need the following results \cite{c27}.
Suppose $k$ is a real number, $0< k\leq1$. Then for any $0\leq x\leq k$,
we have
\begin{equation}\label{n16}
(1+x)^\mu\geq1+\frac{(1+k)^\mu-1}{k^\mu}x^\mu
\end{equation}
for $\mu \geq 1$, and
\begin{equation}\label{n17}
(1+x)^\mu \leq 1+\frac{(1+k)^\mu-1}{k^\mu}x^\mu
\end{equation}
for $0\leq \mu \leq 1$.
Based on the inequality \eqref{n16}, we have the following theorem for R$\alpha$E.

\begin{theorem}
For any multiqubit state $\rho_{AB_0\ldots B_{N-1}}$ and $\alpha\geq2$, we have
\begin{equation}\label{n18}
[E_{\alpha}(\rho_{A|B_0B_1\ldots B_{N-1}})]^\mu
\geq\sum\limits_{j=0}^{N-1}\Big(\frac{(1+k)^\mu-1}{k^\mu}\Big)^{\omega_H(\vec{j})}[E_{\alpha}(\rho_{A|B_j})]^\mu,
\end{equation}
where $\mu\geq1$, $\overrightarrow{j}=\left(j_0, \cdots ,j_{n-1}\right)$ is the vector from the binary representation of $j$, and
$\omega_{H}\left(\overrightarrow{j}\right)$ is the Hamming weight of $\overrightarrow{j}$.
\end{theorem}

{\sf [Proof]} We first prove that
\begin{equation}\label{n19}
\left[\sum\limits_{j=0}^{N-1} E_{\alpha}(\rho_{A|B_j})\right]^\mu
\geq\sum\limits_{j=0}^{N-1}\Big(\frac{(1+k)^\mu-1}{k^\mu}\Big)^{\omega_{H}(\vec{j})}
[E_{\alpha}(\rho_{A|B_j})]^\mu.
\end{equation}
Without loss of generality, we assume that the qubit subsystems $B_0, \ldots, B_{N-1}$ are so labeled such that
\begin{equation}\label{n20}
kE_{\alpha}(\rho_{A|B_j})\geq E_{\alpha}(\rho_{A|B_{j+1}})\geq 0
\end{equation}
for $j=0,1,\ldots,N-2$ and some $0<k\leq1$.

We first show that the inequality \eqref{n19} holds for the case of $N=2^n$.
For $n=1$, let $\rho_{AB_0}$ and $\rho_{AB_1}$ be the two-qubit reduced density matrices
of a three-qubit pure state $\rho_{AB_0 B_1}$. We obtain
\begin{equation}\label{n21}
[E_{\alpha}(\rho_{A|B_0})+E_{\alpha}(\rho_{A|B_1})]^\mu
=[E_{\alpha}(\rho_{A|B_0})]^\mu \Big(1+\frac{E_{\alpha}(\rho_{A|B_1})}{E_{\alpha}(\rho_{A|B_0})}\Big)^\mu.
\end{equation}
Combining \eqref{n16} and \eqref{n20}, we have
\begin{equation}\label{n22}
\Big(1+\frac{E_{\alpha}(\rho_{A|B_1})}{E_{\alpha}(\rho_{A|B_0})}\Big)^\mu \geq
1+\displaystyle\frac{(1+k)^\mu-1}{k^\mu}\Bigg(\frac{E_{\alpha}(\rho_{A|B_1})}{E_{\alpha}(\rho_{A|B_0})}\Bigg)^\mu.
\end{equation}
From \eqref{n21} and \eqref{n22}, we get
\begin{equation*}
[E_{\alpha}(\rho_{A|B_0})+E_{\alpha}(\rho_{A|B_1})]^\mu\geq
[E_{\alpha}(\rho_{A|B_0})]^\mu+\displaystyle\frac{(1+k)^\mu-1}{k^\mu}[E_{\alpha}(\rho_{A|B_1})]^\mu.
\end{equation*}
Therefore, the inequality \eqref{n19} holds for $n=1$.

We assume that the inequality \eqref{n19} holds for $N=2^{n-1}$ with $n\geq 2$,
and prove the case of $N=2^n$.
For an $(N + 1)$-qubit pure state $\rho_{AB_0B_1\ldots B_{N-1}}$,
we have $E_{\alpha}(\rho_{A|B_{j+2^{n-1}}})\leq k^{2^{n-1}}E_{\alpha}(\rho_{A|B_j})$ from \eqref{n20}.
Therefore,
\begin{equation*}
0\leq\frac{\sum\nolimits_{j=2^{n-1}}^{2^n-1}E_{\alpha}(\rho_{A|B_j})}{\sum\nolimits_{j=0}^{2^{n-1}-1}
E_{\alpha}(\rho_{A|B_j})}\leq k^{2^{n-1}}\leq k,
\end{equation*}
and
\begin{equation*}
\Bigg(\sum\nolimits_{j=0}^{N-1}E_{\alpha}(\rho_{A|B_j})\Bigg)^\mu
=\Bigg(\sum\nolimits_{j=0}^{2^{n-1}-1}E_{\alpha}(\rho_{A|B_j})\Bigg)^\mu
\Bigg(1+\frac{\sum_{j=2^{n-1}}^{2^n-1}E_{\alpha}(\rho_{A|B_j})}{\sum_{j=0}^{2^{n-1}-1}E_{\alpha}
(\rho_{A|B_j})}\Bigg)^\mu.
\end{equation*}
Thus, we have
\begin{equation*}
\Bigg(\sum\nolimits_{j=0}^{N-1}E_{\alpha}(\rho_{A|B_j})\Bigg)^\mu
\geq \Bigg(\sum\nolimits_{j=0}^{2^{n-1}-1}E_{\alpha}(\rho_{A|B_j})\Bigg)^\mu
+\displaystyle\frac{(1+k)^\mu-1}{k^\mu}\Bigg(\sum\nolimits_{j=2^{n-1}}^{2^n-1}E_{\alpha}(\rho_{A|B_j})\Bigg)^\mu.
\end{equation*}

According to the induction hypothesis, we get
$$
\Bigg(\sum\nolimits_{j=0}^{2^{n-1}-1}E_{\alpha}(\rho_{A|B_j})\Bigg)^\mu \geq
\sum\nolimits_{j=0}^{2^{n-1}-1}\Big(\frac{(1+k)^\mu-1}{k^\mu}\Big)^{\omega_H(\vec{j})}
[E_{\alpha}(\rho_{A|B_j})]^\mu.
$$
By relabeling the subsystems, the induction hypothesis leads to
$$
\Bigg(\sum\nolimits_{j=2^{n-1}}^{2^n-1}E_{\alpha}(\rho_{A|B_j})\Bigg)^\mu \geq
\sum\nolimits_{j=2^{n-1}}^{2^n-1}\Big(\frac{(1+k)^\mu-1}{k^\mu}\Big)^{\omega_H(\vec{j})-1}[E_{\alpha}(\rho_{A|B_j})]^\mu.
$$
Thus, we have
$$
\Bigg(\sum\nolimits_{j=0}^{2^n-1}E_{\alpha}(\rho_{A|B_j})\Bigg)^\mu\geq
\sum\nolimits_{j=0}^{2^n-1}\Big(\frac{(1+k)^\mu-1}{k^\mu}\Big)^{\omega_H(\vec{j})}[E_{\alpha}(\rho_{A|B_j})]^\mu.
$$

Now consider a $(2^n+1)$-qubit state
\begin{equation}\label{n27}
\Gamma_{AB_0B_1\ldots B_{2^n-1}}=\rho_{AB_0B_1\ldots B_{N-1}}\otimes \sigma_{B_N\ldots B_{2^n-1}},
\end{equation}
which is the tensor product of $\rho_{AB_0B_1\ldots B_{N-1}}$ and an arbitrary $(2^n-N)$-qubit state $\sigma_{B_N\ldots B_{2^n-1}}$. We have
\begin{equation*}
[E_{\alpha}(\Gamma_{A|B_0B_1\ldots B_{2^n-1}})]^\mu
\geq\sum\nolimits_{j=0}^{2^n-1}\Big(\frac{(1+k)^\mu-1}{k^\mu}\Big)^{\omega_H(\vec{j})}[E_{\alpha}(\Gamma_{A|B_j})]^\mu,
\end{equation*}
where $\Gamma_{A|B_j}$ is the two-qubit reduced density matrix of $\Gamma_{AB_0B_1\ldots B_{2^n-1}}$, $j=0,1,\ldots,2^n-1$. Therefore,
\begin{align*}
[E_{\alpha}(\rho_{A|B_0B_1\ldots B_{N-1}})]^\mu =&[E_{\alpha}(\Gamma_{A|B_0B_1\ldots B_{2^n-1}})]^\mu \nonumber\\
\geq& \sum\nolimits_{j=0}^{2^n-1}\Big(\frac{(1+k)^\mu-1}{k^\mu}\Big)^{\omega_H(\vec{j})}[E_{\alpha}(\Gamma_{A|B_j})]^\mu \nonumber\\
=& \sum\nolimits_{j=0}^{N-1}\Big(\frac{(1+k)^\mu-1}{k^\mu}\Big)^{\omega_H(\vec{j})}[E_{\alpha}(\rho_{A|B_j})]^\mu,
\end{align*}
where $\Gamma_{A|B_0B_1\ldots B_{2^n-1}}$ is separated to the bipartition $AB_0\ldots B_{N-1}$ and $B_N\ldots B_{2^n-1}$, $E_{\alpha}\left(\Gamma_{A|B_0 B_1 \cdots B_{2^n-1}}\right)=E_{\alpha}\left(\rho_{A|B_0 B_1 \cdots B_{N-1}}\right)$,
$E_{\alpha}\left(\Gamma_{A|B_j}\right)=0$ for $j=N, \cdots , 2^n-1$,
and $\Gamma_{AB_j}=\rho_{AB_j}$ for each $j=0, \cdots , N-1$.
\qed

Since $\Big(\frac{(1+k)^\mu-1}{k^\mu}\Big)^{\omega_H(\vec{j})}\geq \mu^{\omega_H(\vec{j})}$ for $\mu \geq1$, for any multiqubit state $\rho_{AB_0 B_1 \cdots B_{N-1}}$ we have the following relation,
\begin{equation*}
[E_{\alpha}(\rho_{A|B_0B_1\ldots B_{N-1}})]^\mu
\geq\sum\nolimits_{j=0}^{N-1}\Big(\frac{(1+k)^\mu-1}{k^\mu}\Big)^{\omega_H(\vec{j})}[E_{\alpha}(\rho_{A|B_j})]^\mu
\geq\sum\nolimits_{j=0}^{N-1}\mu^{\omega_H(\vec{j})}[E_{\alpha}(\rho_{A|B_j})]^\mu.
\end{equation*}
Therefore, our inequality \eqref{n18} in Theorem 1 is always tighter than the inequality \eqref{n12}.

In fact, the tighter monogamy inequality \eqref{n18} holds not only for multiqubit systems, but also for
some multipartite higher-dimensional quantum systems, which can be proved in a similar way as in \cite{c25}. Here, we show that \eqref{n18} is also more efficient than \eqref{n12} for such higher-dimensional quantum systems. Let us consider the counterexample of the CKW inequality in tripartite quantum systems \cite{nc30},
\begin{equation}\label{psi}
|\psi\rangle_{ABC}=\frac{1}{\sqrt{6}}(|123\rangle-|132\rangle+|231\rangle-|213\rangle+|312\rangle-|321\rangle).
\end{equation}
One has $E_{\alpha}(|\psi\rangle_{A|BC})=S_\alpha(\rho)$.
Taking $\alpha=3$, we have $E_{\alpha}(|\psi\rangle_{A|BC})=\log 3$
and the R$\alpha$E of the two-qubit reduced density matrices are
\begin{equation*}
E_{\alpha}(\rho_{A|B})=E_{\alpha}(\rho_{A|C})=-\frac{1}{2}\log \mathrm{tr}\sigma^{3}_A=1.
\end{equation*}
In this case $k=1$, for $\mu\geq1$, we have
\begin{equation*}
y_1\equiv [E_{\alpha}(\rho_{A|B})]^\mu+\frac{(1+k)^\mu-1}{k^\mu}[E_{\alpha}(\rho_{A|C})]^\mu
=1+\frac{(1+k)^\mu-1}{k^\mu}
=2^\mu,
\end{equation*}
and
\begin{equation*}
y_2\equiv[E_{\alpha}(\rho_{A|B})]^\mu+\mu[E_{\alpha}(\rho_{A|C})]^\mu=1+\mu.
\end{equation*}
Therefore, one gets
\begin{equation*}
[E_{\alpha}(\rho_{A|B})]^\mu+\frac{(1+k)^\mu-1}{k^\mu}[E_{\alpha}(\rho_{A|C})]^\mu \geq [E_{\alpha}(\rho_{A|B})]^\mu+\mu[E_{\alpha}(\rho_{A|C})]^\mu.
\end{equation*}
where $\mu\geq1$, see Fig. 1. In other words, our new monogamy inequality is indeed tighter than the previous one given in \cite{c25}.

\begin{figure}
  \centering
  \includegraphics[width=7cm]{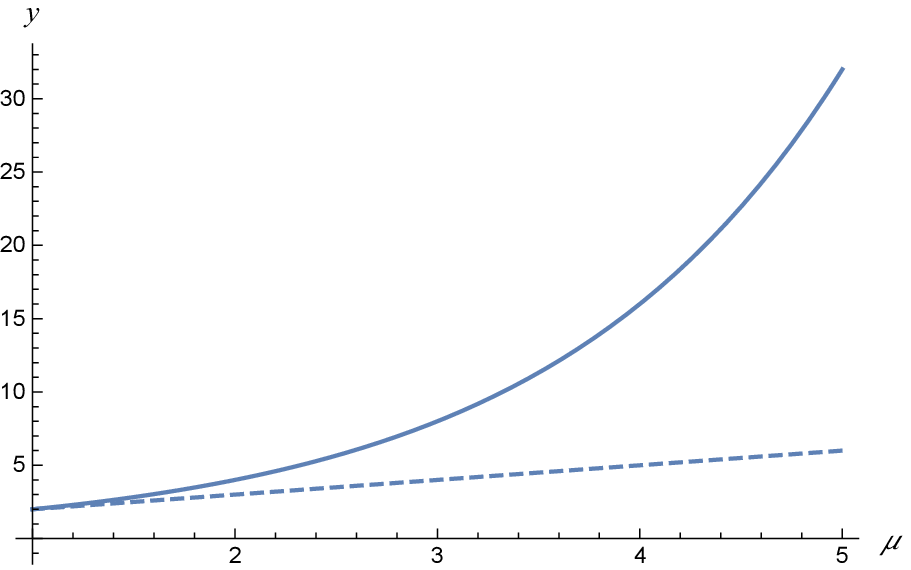}\\
  \caption{R\'{e}nyi-$\alpha$ entanglement with respect to $\mu$:
  the solid line is for $y_1$ and the dashed line for $y_2$ from the result in \cite{c25}.}
\end{figure}

Under certain conditions, the inequality \eqref{n18} can even be improved
further to become a much tighter inequality.

\begin{theorem}
For $\mu\geq1$, $\alpha\geq2$ and real number $0< k \leq1$, any multiqubit state $\rho_{AB_0\ldots B_{N-1}}$ satisfies
\begin{equation}\label{n31}
[E_{\alpha}(\rho_{A|B_0\ldots B_{N-1}})]^\mu
\geq\sum\nolimits_{j=0}^{N-1}\Big(\frac{(1+k)^\mu-1}{k^\mu}\Big)^j[E_{\alpha}(\rho_{A|B_j})]^\mu,
\end{equation}
if
\begin{equation}\label{n32}
kE_{\alpha}(\rho_{A|B_j})\geq\sum\nolimits_{j=i+1}^{N-1}E_{\alpha}(\rho_{A|B_j})
\end{equation}
for $i=0,1,\ldots,N-2$.
\end{theorem}

{\sf [Proof] }
We need to show
\begin{equation}
\left(\sum_{j=0}^{N-1}E_{\alpha}\left(\rho_{A|B_j}\right)\right)^{\mu}
\geq \sum_{j=0}^{N-1} \left( \frac{(1+k)^\mu-1}{k^\mu}\right)^{j}\left(E_{\alpha}\left(\rho_{A|B_j}\right)\right)^{\mu}.
\end{equation}

For any multiqubit state $\rho_{AB_0\ldots B_{N-1}}$,
it is easy to show that
\begin{equation*}
\left(\sum_{j=0}^{N-1}E_{\alpha}\left(\rho_{A|B_j}\right)\right)^{\mu}
=\left(E_{\alpha}\left(\rho_{A|B_0}\right)\right)^{\mu}
\left(1+\frac{\sum_{j=1}^{N-1}E_{\alpha}\left(\rho_{A|B_j}\right)}
{E_{\alpha}\left(\rho_{A|B_0}\right)} \right)^{\mu}
\end{equation*}
and
\begin{equation*}
\left(1+\frac{\sum_{j=1}^{N-1}E_{\alpha}\left(\rho_{A|B_j}\right)}
{E_{\alpha}\left(\rho_{A|B_0}\right)} \right)^{\mu}
\geq 1 + \left( \frac{(1+k)^\mu-1}{k^\mu}\right) \left(\frac{\sum_{j=1}^{N-1}E_{\alpha}\left(\rho_{A|B_j}\right)}
{E_{\alpha}\left(\rho_{A|B_0}\right)}\right)^{\mu}.
\end{equation*}
Thus,
\begin{align*}
\left(\sum_{j=0}^{N-1}E_{\alpha}\left(\rho_{A|B_j}\right) \right)^{\mu}
\geq & \left(E_{\alpha}\left(\rho_{A|B_0}\right)\right)^{\mu}
+\left( \frac{(1+k)^\mu-1}{k^\mu}\right) \left(\sum_{j=1}^{N-1}E_{\alpha}\left(\rho_{A|B_j}\right) \right)^{\mu} \nonumber\\
\geq & \sum_{j=0}^{N-1} \left( \frac{(1+k)^\mu-1}{k^\mu}\right)^{j}\left(E_{\alpha}\left(\rho_{A|B_j}\right)\right)^{\mu},
\end{align*}
where the second inequality is due to the induction hypothesis.
\qed

In fact, according to \eqref {n11}, for any $\mu \geq 1$, one has
\begin{align*}
[E_{\alpha}(\rho_{A|B_0\ldots B_{N-1}})]^\mu
\geq & \sum_{j=0}^{N-1} \left( \frac{(1+k)^\mu-1}{k^\mu}\right)^{j}\left(E_{\alpha}\left(\rho_{A|B_j}\right)\right)^{\mu} \nonumber\\
\geq & \sum\nolimits_{j=0}^{N-1}\Big(\frac{(1+k)^\mu-1}{k^\mu}\Big)^{\omega_H(\vec{j})}[E_{\alpha}(\rho_{A|B_j})]^\mu.
\end{align*}

For the case of $\mu<0$, we can also derive a tighter upper bound of
$E^\mu_{\alpha}(\rho_{A|B_0B_1\ldots B_{N-1}})$.

\begin{theorem}
For any multiqubit state $\rho_{AB_0\ldots B_{N-1}}$ with $E_{\alpha}(\rho_{AB_i})\neq0$,
$i=0,1,\ldots,N-1$,
we have
\begin{equation}\label{SC28}
[E_{\alpha}(\rho_{A|B_0B_1\ldots B_{N-1}})]^\mu
\leq\frac{1}{N-1}\sum\nolimits_{j=0}^{N-1}[E_{\alpha}(\rho_{A|B_j})]^\mu,
\end{equation}
for all $\mu<0$ and $\alpha\geq2$.
\end{theorem}

{\sf [Proof] }
Similar to the proof in \cite{jin}, for arbitrary three-qubit states we have
\begin{align}\label{SCREN1}
[E_{\alpha}(\rho_{A|B_0B_1})]^\mu
\leq & [E^{2}_{\alpha}(\rho_{A|B_0})+ E^{2}_{\alpha}(\rho_{A|B_1})]^{\frac{\mu}{2}}\nonumber\\
=& [E_{\alpha}(\rho_{A|B_0})]^\mu \Big(1+\frac{E^{2}_{\alpha}(\rho_{A|B_1})}{E^{2}_{\alpha}(\rho_{A|B_0})}\Big)^{\frac{\mu}{2}}\nonumber\\
<& [E_{\alpha}(\rho_{A|B_0})]^\mu,
\end{align}
where the first inequality is from $\mu<0$, the second inequality is due to
$\Big(1+\frac{E^{2}_{\alpha}(\rho_{A|B_1})}{E^{2}_{\alpha}(\rho_{A|B_0})}\Big)^\mu<1$.
Moreover, we have
\begin{equation}\label{SCREN2}
[E_{\alpha}(\rho_{A|B_0B_1})]^\mu<[E_{\alpha}(\rho_{A|B_1})]^\mu.
\end{equation}
Combining \eqref{SCREN1} and \eqref{SCREN2}, we get
\begin{equation*}
[E_{\alpha}(\rho_{A|B_0B_1})]^\mu
<\frac{1}{2}\{[E_{\alpha}(\rho_{A|B_0})]^\mu+[E_{\alpha}(\rho_{A|B_1})]^\mu\}.
\end{equation*}
Thus, we obtain
\begin{align}\label{S23}
&[E_{\alpha}(\rho_{A|B_0B_1\ldots B_{N-1}})]^\mu \nonumber\\
< & \frac{1}{2}\Bigg\{\Big[E_{\alpha}(\rho_{A|B_0})\Big]^\mu+\Big[E_{\alpha}(\rho_{A|B_1\ldots B_{N-1}})\Big]^\mu\Bigg\}\nonumber\\
<& \frac{1}{2}\Big[E_{\alpha}(\rho_{A|B_0})\Big]^\mu+\Big(\frac{1}{2}\Big)^2\Big[E_{\alpha}(\rho_{A|B_1})\Big]^\mu
+\Big(\frac{1}{2}\Big)^2\Big[E_{\alpha}(\rho_{A|B_2\ldots B_{N-1}})\Big]^\mu \nonumber\\
<& \ldots \nonumber\\
<& \frac{1}{2}\Big[E_{\alpha}(\rho_{A|B_0})\Big]^\mu+\Big(\frac{1}{2}\Big)^2\Big[E_{\alpha}(\rho_{A|B_1})\Big]^\mu+\ldots
+\Big(\frac{1}{2}\Big)^{N-2}\Big[E_{\alpha}(\rho_{A|B_{N-2}})\Big]^\mu
+\Big(\frac{1}{2}\Big)^{N-2}\Big[E_{\alpha}(\rho_{A|B_{N-1}})\Big]^\mu.
\end{align}
One can get a set of inequalities through the cyclic permutation of the pair indices $B_0$, $B_1,$ $\ldots$, $B_{N-1}$ in \eqref {S23}. Summing up these inequalities, we get \eqref {SC28}.
\qed

\section{Tighter constraints of multiqubit entanglement in terms of R$\alpha$EoA}

We consider now the R\'{e}nyi-$\alpha$ entanglement of assistance (R$\alpha$EoA) defined in (\ref{EoA}),
and provide a class of polygamy inequalities satisfied by the multiqubit entanglement in terms of R$\alpha$EoA.

\begin{theorem}
For any multiqubit state $\rho_{AB_0\ldots B_{N-1}}$ and $0\leq \mu \leq1$, $0<\alpha<2$, $\alpha\neq1$,
we have
\begin{equation}\label{n38}
[E^{a}_{\alpha}(\rho_{A|B_0B_1\ldots B_{N-1}})]^\mu
\leq\sum\limits_{j=0}^{N-1}\Big(\frac{(1+k)^\mu-1}{k^\mu}\Big)^{\omega_H(\vec{j})}[E^{a}_{\alpha}(\rho_{A|B_j})]^\mu.
\end{equation}
\end{theorem}

{\sf [Proof] }
Similar to proof in Ref. \cite{c25},
we just need to prove
\begin{equation}\label{n39}
\left[\sum\limits_{j=0}^{N-1} E^{a}_{\alpha}(\rho_{A|B_j})\right]^\mu
\leq\sum\limits_{j=0}^{N-1}\Big(\frac{(1+k)^\mu-1}{k^\mu}\Big)^{\omega_{H}(\vec{j})}
[E^{a}_{\alpha}(\rho_{A|B_j})]^\mu.
\end{equation}
Firstly, assume that the qubit subsystems $B_0, \ldots, B_{N-1}$ satisfies
\begin{equation}\label{n40}
kE^{a}_{\alpha}(\rho_{A|B_j})\geq E^{a}_{\alpha}(\rho_{A|B_{j+1}})\geq 0,
\end{equation}
where $j=0,1,\ldots,N-2$ and $0<k\leq1$.
Similar to the proof of Theorem 1, we first show that the inequality \eqref{n39} holds for a three-qubit pure state $\rho_{AB_0 B_1}$. We have
\begin{align*}
[E^{a}_{\alpha}(\rho_{A|B_0})+E^{a}_{\alpha}(\rho_{A|B_1})]^\mu
=& [E^{a}_{\alpha}(\rho_{A|B_0})]^\mu \Big(1+\frac{E^{a}_{\alpha}(\rho_{A|B_1})}{E^{a}_{\alpha}(\rho_{A|B_0})}\Big)^\mu \nonumber\\
\leq & [E^{a}_{\alpha}(\rho_{A|B_0})]^\mu \left[1+\displaystyle\frac{(1+k)^\mu-1}{k^\mu}\Bigg(\frac{E^{a}_{\alpha}(\rho_{A|B_1})}{E^{a}_{\alpha}(\rho_{A|B_0})}\Bigg)^\mu \right] \nonumber\\
=& [E^{a}_{\alpha}(\rho_{A|B_0})]^\mu+\displaystyle\frac{(1+k)^\mu-1}{k^\mu}[E^{a}_{\alpha}(\rho_{A|B_1})]^\mu,
\end{align*}
where the inequality is due to \eqref {n17}.

Then we assume that the inequality \eqref {n39} holds for $N=2^{n-1}$ with $n\geq 2$.
Consider the case of $N=2^n$.
For an $(N + 1)$-qubit pure state $\rho_{AB_0B_1\ldots B_{N-1}}$ with its two-qubit reduced density matrices $\rho_{AB_j}$, $j=0,1,\ldots,N-1$, we have $E^{a}_{\alpha}(\rho_{A|B_{j+2^{n-1}}})\leq k^{2^{n-1}}E^{a}_{\alpha}(\rho_{A|B_j})$ due to the ordering of subsystems in the inequality \eqref{n40}.
Then, we get
\begin{equation*}
0\leq\frac{\sum\nolimits_{j=2^{n-1}}^{2^n-1}E^{a}_{\alpha}(\rho_{A|B_j})}{\sum\nolimits_{j=0}^{2^{n-1}-1}
E^{a}_{\alpha}(\rho_{A|B_j})}\leq k^{2^{n-1}}\leq k\leq 1,
\end{equation*}
and
\begin{equation*}
\Bigg(\sum\nolimits_{j=0}^{N-1}E^{a}_{\alpha}(\rho_{A|B_j})\Bigg)^\mu
=\Bigg(\sum\nolimits_{j=0}^{2^{n-1}-1}E^{a}_{\alpha}(\rho_{A|B_j})\Bigg)^\mu
\Bigg(1+\frac{\sum_{j=2^{n-1}}^{2^n-1}E^{a}_{\alpha}(\rho_{A|B_j})}{\sum_{j=0}^{2^{n-1}-1}E^{a}_{\alpha}
(\rho_{A|B_j})}\Bigg)^\mu.
\end{equation*}
Hence,
\begin{equation*}
\Bigg(\sum\nolimits_{j=0}^{N-1}E^{a}_{\alpha}(\rho_{A|B_j})\Bigg)^\mu
\leq \Bigg(\sum\nolimits_{j=0}^{2^{n-1}-1}E_{\alpha}(\rho_{A|B_j})\Bigg)^\mu
+\displaystyle\frac{(1+k)^\mu-1}{k^\mu}\Bigg(\sum\nolimits_{j=2^{n-1}}^{2^n-1}E^{a}_{\alpha}(\rho_{A|B_j})\Bigg)^\mu.
\end{equation*}
According to the induction hypothesis, we get
$$
\Bigg(\sum\nolimits_{j=0}^{2^{n-1}-1}E^{a}_{\alpha}(\rho_{A|B_j})\Bigg)^\mu
\leq \sum\nolimits_{j=0}^{2^{n-1}-1}\Big(\frac{(1+k)^\mu-1}{k^\mu}\Big)^{\omega_H(\vec{j})}
[E^{a}_{\alpha}(\rho_{A|B_j})]^\mu.
$$
By relabeling the subsystems, the induction hypothesis leads to
$$
\Bigg(\sum\nolimits_{j=2^{n-1}}^{2^n-1}E^{a}_{\alpha}(\rho_{A|B_j})\Bigg)^\mu
\leq \sum\nolimits_{j=2^{n-1}}^{2^n-1}\Big(\frac{(1+k)^\mu-1}{k^\mu}\Big)^{\omega_H(\vec{j})-1}[E^{a}_{\alpha}(\rho_{A|B_j})]^\mu.
$$
Therefore,
$$
\Bigg(\sum\nolimits_{j=0}^{2^n-1}E^{a}_{\alpha}(\rho_{A|B_j})\Bigg)^\mu
\leq \sum\nolimits_{j=0}^{2^n-1}\Big(\frac{(1+k)^\mu-1}{k^\mu}\Big)^{\omega_H(\vec{j})}[E^{a}_{\alpha}(\rho_{A|B_j})]^\mu.
$$

Consider the $(2^n+1)$-qubit state \eqref{n27}. We have
\begin{align*}
[E^{a}_{\alpha}(\rho_{A|B_0B_1\ldots B_{N-1}})]^\mu =&[E^{a}_{\alpha}(\Gamma_{A|B_0B_1\ldots B_{2^n-1}})]^\mu \nonumber\\
\leq & \sum\nolimits_{j=0}^{2^n-1}\Big(\frac{(1+k)^\mu-1}{k^\mu}\Big)^{\omega_H(\vec{j})}[E^{a}_{\alpha}(\Gamma_{A|B_j})]^\mu \nonumber\\
=& \sum\nolimits_{j=0}^{N-1}\Big(\frac{(1+k)^\mu-1}{k^\mu}\Big)^{\omega_H(\vec{j})}[E^{a}_{\alpha}(\rho_{A|B_j})]^\mu.
\end{align*}
\qed

Since $\frac{(1+k)^\mu-1}{k^\mu}\leq\mu$ for $0\leq \mu \leq1$, it is easy to see that \eqref{n38} is tighter than \eqref{n13}.

As an example, let us consider the three-qubit W-state \cite{c31},
\begin{equation}\label{psi}
|W\rangle_{ABC}=\frac{1}{\sqrt{3}}(|211\rangle+|121\rangle+|112\rangle).
\end{equation}
We have $E^{a}_{\alpha}(|\psi\rangle_{A|BC})=S_\alpha(\rho)=\log 3-\frac{2}{3}$
and
\begin{equation*}
E^{a}_{\alpha}(\rho_{A|B})=E^{a}_{\alpha}(\rho_{A|C})=-\frac{1}{2}\log \mathrm{tr}\sigma^{3}_A=\frac{2}{3}.
\end{equation*}
In the case of $k=1$ and $0\leq \mu \leq1$, we have
\begin{equation*}
y_3 \equiv [E^{a}_{\alpha}(\rho_{A|B})]^\mu+\frac{(1+k)^\mu-1}{k^\mu}[E^{a}_{\alpha}(\rho_{A|C})]^\mu
=(\frac{2}{3})^{\mu}+\frac{(1+k)^\mu-1}{k^\mu}(\frac{2}{3})^{\mu}
=(\frac{2}{3})^{\mu}2^{\mu},
\end{equation*}
and
\begin{equation*}
y_4 \equiv [E^{a}_{\alpha}(\rho_{A|B})]^\mu+\mu[E^{a}_{\alpha}(\rho_{A|C})]^\mu=(\frac{2}{3})^{\mu}(1+\mu).
\end{equation*}

Therefore, we get
\begin{equation*}
[E^{a}_{\alpha}(\rho_{A|B})]^\mu+\frac{(1+k)^\mu-1}{k^\mu}[E^{a}_{\alpha}(\rho_{A|C})]^\mu \leq [E_{\alpha}(\rho_{A|B})]^\mu+\mu[E_{\alpha}(\rho_{A|C})]^\mu
\end{equation*}
where $0\leq \mu\leq1$, see Fig. 2.

\begin{figure}
\centering
\includegraphics[width=7cm]{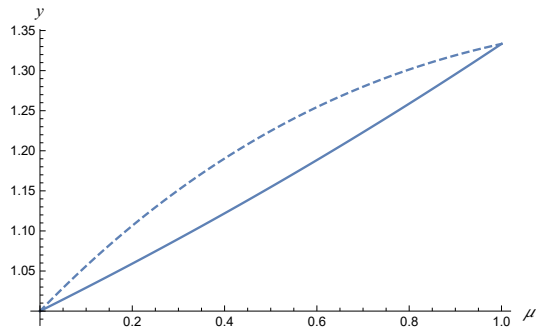}\\
\caption{R\'{e}nyi-$\alpha$ entanglement $y$ with respect to $\mu$:
the solid line is for $y_3$ and the dashed line for $y_4$ from the result in \cite{c25}.}
\end{figure}

Similar to the improvement from the inequality \eqref {n18} to the inequality \eqref{n31},
we can also improve the polygamy inequality in Theorem 4. The proof is similar to the Theorem 2.

\begin{theorem}
For $0 \leq \mu \leq 1$, $0<\alpha<2$, $\alpha \neq1$ and  $0< k \leq1$,
we have for any multiqubit state $\rho_{AB_0\ldots B_{N-1}}$,
\begin{equation}\label{ps}
[E^{a}_{\alpha}(\rho_{A|B_0\ldots B_{N-1}})]^\mu
\leq \sum\nolimits_{j=0}^{N-1}\Big(\frac{(1+k)^\mu-1}{k^\mu}\Big)^j[E^{a}_{\alpha}(\rho_{A|B_j})]^\mu,
\end{equation}
if
\begin{equation}\label{psi1}
kE^{a}_{\alpha}(\rho_{A|B_i})\geq\sum\nolimits_{j=i+1}^{N-1}E^{a}_{\alpha}(\rho_{A|B_j})
\end{equation}
for $i=0,1,\ldots,N-2$.
\end{theorem}

Since $\omega_H(\vec{j})\leq j$, for $0\leq\mu\leq1$ we obtain
\begin{align*}
[E^{a}_{\alpha}(\rho_{A|B_0\ldots B_{N-1}})]^\mu
\leq & \sum_{j=0}^{N-1} \left( \frac{(1+k)^\mu-1}{k^\mu}\right)^{j}\left(E^{a}_{\alpha}\left(\rho_{A|B_j}\right)\right)^{\mu} \nonumber\\
\leq & \sum\nolimits_{j=0}^{N-1}\Big(\frac{(1+k)^\mu-1}{k^\mu}\Big)^{\omega_H(\vec{j})}[E^{a}_{\alpha}(\rho_{A|B_j})]^\mu.
\end{align*}
Therefore, for any multiqubit state $\rho_{AB_0\ldots B_{N-1}}$ satisfying the condition \eqref {psi1}, the inequality \eqref {ps} of Theorem 5 is tighter than the inequality \eqref {n38} of Theorem 4.

\section{conclusion}
Quantum entanglement is the essential resource in quantum information.
The monogamy and polygamy relations characterize the entanglement distributions
in the multipartite systems. Tighter monogamy and polygamy inequalities give finer
characterization of the entanglement distribution.
In this article, by using the Hamming weights of binary vectors we have proposed a class of monogamy inequalities related to the $\mu$th power of the entanglement measure based on R\'{e}nyi-$\alpha$ entropy, polygamy relations in terms of the $\mu$th powered of of R$\alpha$EoA for $0\leq \mu \leq 1$.
These new monogamy and polygamy relations are shown to be tighter than the existing ones.
Moreover, it has been shown that our monogamy inequality is effective for the counterexamples
of the CKW monogamy inequality in higher-dimensional systems.
Our results may highlight further investigations on the entanglement distribution in multipartite systems.

\bigskip
\noindent{\bf Acknowledgments}\, \, This work is supported by NSFC under numbers 11765016, 11675113, NSF of Beijing under No. KZ201810028042, and Beijing Natural Science Foundation (Z190005).

\end{document}